\documentclass[aoas,nameyear,rotating,secthm,seceqn,dvips]{arximspdf}
\usepackage{multirow,dcolumn}
\usepackage{graphicx}

\doi{10.1214/09-AOAS323}
\volume{4}
\issue{3}
\pubyear{2010}
\firstpage{1214}
\lastpage{1233}

\makeatletter
\newproclaim{remark}{Remark}
\newcolumntype{d}[1]{D{.}{.}{#1}}

  \let\sv@tabnotetext\tabnotetext
  \let\sv@tabnotemark@fmt\tabnotemark@fmt
   \long\def\legend#1{{\let\tabnote@indent\leavevmode\sv@tabnotetext[]{}{#1}}}

\makeatother

\begin{document}
\begin{frontmatter}

\title{A general statistical framework for dissecting parent-of-origin
effects underlying endosperm traits in flowering plants\protect\thanksref{T1}}
\pdftitle{A general statistical framework for dissecting parent-of-origin
effects underlying endosperm traits in flowering plants}
\runtitle{Mapping endosperm \textsc{iQTL}'s with variance components model}

\thankstext{T1}{Supported in part by NSF Grant DMS-0707031 and by
Michigan State University intramural research Grant 06-IRGP-789.}

\begin{aug}
\author[A]{\fnms{Gengxin} \snm{Li}\ead[label=e1]{ligengxi@stt.msu.edu}}
and
\author[A]{\fnms{Yuehua} \snm{Cui}\ead[label=e2]{cui@stt.msu.edu}\corref{}}

\affiliation{Michigan State University}

\runauthor{G. Li and Y. Cui}

\address[A]{Department of Statistics and Probability\\
Michigan State University\\
East Lansing, Michigan 48824\\USA\\
\printead{e1}\\
\hphantom{E-mail: }\printead*{e2}}
\end{aug}

% HISTORY:
\received{\smonth{8} \syear{2009}}
\revised{\smonth{12} \syear{2009}}

% ABSTRACT
%
\begin{abstract}
Genomic imprinting has been thought to play an important role in seed
development in flowering plants. Seed in a flowering plant normally
contains diploid embryo and triploid endosperm. Empirical studies have
shown that some economically important endosperm traits are genetically
controlled by imprinted genes. However, the exact number and location
of the imprinted genes are largely unknown due to the lack of efficient
statistical mapping methods. Here we propose a general statistical
variance components framework by utilizing the natural information of
sex-specific allelic sharing among sibpairs in line crosses, to map
imprinted quantitative trait loci (iQTL) underlying endosperm traits.
We propose a new variance components partition method considering the
unique characteristic of the triploid endosperm genome, and develop a
restricted maximum likelihood estimation method in an interval scan for
estimating and testing genome-wide iQTL effects. Cytoplasmic maternal
effect which is thought to have primary influences on yield and grain
quality is also considered when testing for genomic imprinting.
Extension to multiple iQTL analysis is proposed. Asymptotic
distribution of the likelihood ratio test for testing the variance
components under irregular conditions are studied. Both simulation
study and real data analysis indicate good performance and powerfulness
of the developed approach.
\end{abstract}

% KEYWORDS
%

\end{frontmatter}

%s1 ###
\section{Introduction}\label{sec1}
The life cycle of an angiosperm starts with the process of double
fertilization, where the fertilization of the haploid egg with one
sperm cell forms the embryo, and the fusion of the two polar nuclei
with another sperm cell develops into endosperm [Chaudhury et al.
(\citeyear{Cetal2001})]. Thus, endosperm is a tissue unique to angiosperm. The embryo
and endosperm are genetically identical, except that the endosperm is
triploid composed of one set of paternal and two identical sets of
maternal chromosomes. In cereals, the endosperm of a grain is the major
storage organ providing nutrition for early-stage seed development, and
more than that, serves as the major source of food for human beings.
The identification of important genes that underlie the variation of
quantitative traits of various interests in endosperm is thus
paramountly important.

Genomic imprinting refers to the situation where the expression of the
same genes is different depending on their parental origin [Pfifer
(\citeyear{P2000})]. It has been increasingly recognized that many endosperm traits
are controlled by genomic imprinting. For example, endoreduplication is
a commonly observed phenomenon which shows a maternally controlled
parent-of-origin effect in maize endosperm [Dilkes et al. (\citeyear{Detal2002})].
Cells undergoing endoreduplication are typically larger than other
cells, which consequently results in larger fruits or seeds beneficial
to human beings [Grime and Mowforth (\citeyear{GM1982})]. Other reports of genomic
imprinting with paternal imprinting in maize endosperm include, for
instance, the $r$ gene in the regulation of anthocyanin [Kermicle
(\citeyear{K1970})], the seed storage protein regulatory gene \textit{dsrl} [Chaudhuri
and Messing (\citeyear{CM1994})], the \textit{MEA} gene affecting seed development
[Kinoshita et al. (\citeyear{Ketal1999})] and some $\alpha$-\textit{tubulin} genes
[Lund, Messing and Viotti (\citeyear{LMV1995})]. These studies underscore the value of developing
statistical methods that empower geneticists to identify the
distribution and effects of imprinted genes controlling endosperm
traits.

Statistical methods for mapping imprinted genes or imprinted
quantitative trait loci (iQTL) have been extensively studied. Focusing
on different genetic designs and different segregation populations,
methods were developed in mapping iQTL underlying quantitative traits
in controlled experimental crosses [e.g., Cui, Cheverud and Wu (\citeyear{CCW2007}); Cui et al. (\citeyear{Cetal2006}); Wolf et al.
(\citeyear{Wetal2008})], in outbred population [e.g., de Koning, Bovenhuis and van Arendonk (\citeyear{KBA2002})] and in human
population [e.g., Hanson et al. (\citeyear{Hetal2001}); Shete, Zhou and
Amos (\citeyear{SZA2003})]. Broadly speaking, these methods can be
categorized into two frameworks: one based on the fixed effect model
where the iQTL effect is considered as fixed [e.g., Cui et al. (\citeyear{Cetal2006}, \citeyear{Cetal2007});
de Koning, Bovenhuis and van Arendonk (\citeyear{KBA2002})], and the other considering iQTL effect
as random and estimating the genetic variances contributed by an iQTL
[e.g., Hanson et al. (\citeyear{Hetal2001}); Shete, Zhou and Amos (\citeyear{SZA2003}); Li and Cui (\citeyear{LC2009a})].
The method proposed by Li and Cui (\citeyear{LC2009a}) extended the variance
components model to experimental crosses and showed relative merits in
mapping iQTLs with inbred lines. However, all these approaches for iQTL
mapping were developed based on diploid populations, whereby
chromosomes are paired. Their applications are immediately limited when
the ploidy level of the study population is more than two, for
instance, the triploid endosperm.

In this study we propose to extend our previous work in iQTL mapping
with the variance components approach in experimental crosses [Li and
Cui (\citeyear{LC2009a})], and consider the unique genetic makeup of
the triploid endosperm genome to map iQTLs underlying triploid
endosperm traits. Cytoplasmic maternal effects are also considered and
adjusted when testing for genomic imprinting. Motivated by a real
experiment, we propose a reciprocal backcross design initiated with two
inbred lines. The likelihood ratio test (LRT) is applied to test the
significance of the variance components and its asymptotic distribution
is evaluated under irregular conditions.

The article is organized as follows. Section \ref{sec2} will illustrate the
basic genetic design and the statistical mapping framework. We propose
a new approach for calculating the parental specific allelic sharing
among inbreeding triploid sibs. Statistical hypothesis testings are
proposed to assess iQTL effects. The limiting distribution of the LRT
under the proposed mapping framework is studied. The multiple iQTL
model is also proposed to separate closely linked (i)QTLs. Sections \ref{sec3}
and \ref{sec4} will be devoted to simulations and real application followed by a
general discussion in Section \ref{sec5}.

%s2 ###
\section{Statistical method}\label{sec2}
%s2.1 ###
\subsection{The genetic design}\label{sec2.1}
Using experimental crosses for QTL mapping has been the traditional
means in targeting genetic regions harboring potential genes
responsible for quantitative trait variations. Toward the goal of
mapping iQTL underlying endosperm traits in line crosses, we propose a
reciprocal backcross design. A similar design was proposed by Li and
Cui (\citeyear{LC2009a}) for diploid mapping populations. In brief, two
inbred parents with genotypes $AA$ and $aa$ are crossed to produce an
F$_1$ population ($Aa$). F$_1$ individuals are then backcrossed with
one of the parents to generate backcross populations. We can use both
parents as the maternal strain to cross with an F$_1$ individual to
generate two backcross segregation populations. Or we can use F$_1$
individuals as the maternal strains to cross with both parents to
produce another two sets of segregation populations. The so-called
reciprocal backcross design generates four different segregation
populations with each one being considered as one family. Large number
of backcross families can be obtained by simply replicating each one of
the above crosses.

To distinguish the allelic parental origin, we use subscript letters
$f$ and $m$ to denote an allele inherited from the father and mother,
respectively. A list of possible offspring genotypes considering the
unique genetic makeups in the triploid endosperm genome is detailed in
the second column in Table \ref{IBD}. Clearly, the endosperm genome
carries one extra maternal copy due to the unique double fertilization
step in flowering plants. When a dosage effect is considered, we do
expect different expression values triggered by endosperm and embryo
genes.

%t1 ###
\begin{sidewaystable}
\tablewidth=\textheight
\tablewidth=\textwidth
\caption{The allelic-specific IBD sharing coefficients for full-sib pairs
in a reciprocal backcross design}\label{IBD}
\begin{tabular*}{\textwidth}{@{\extracolsep{\fill}}lcccccccccccc@{}}
\hline
&\multirow{2}{38pt}[-6pt]{\centering\textbf{Offspring genotype}}&\multicolumn{9}{c}{\textbf{Parent-specific IBD sharing}}&\multicolumn{2}{c@{}}{\textbf{Total IBD}}\\[-5pt]
&&\multicolumn{9}{c}{\hrulefill}&\multicolumn{2}{c@{}}{\hrulefill}\\
\textbf{Backcross}&&\multicolumn{2}{c}{$\bolds{\pi_{mm}}$}&&\multicolumn{2}{c}{$\bolds{\pi_{ff}}$}&&
\multicolumn{2}{c}{$\bolds{\pi_{m/f}}$}&&\multicolumn{2}{c@{}}{$\bolds{\pi}$}\\
\hline
&&$Q_mQ_mQ_f$&$Q_mQ_mq_f$&&$Q_mQ_mQ_f$&$Q_mQ_mq_f$&&$Q_mQ_mQ_f$&$Q_mQ_mq_f$ &&$Q_mQ_mQ_f$&$Q_mQ_mq_f$\\
$QQ\times Qq$&$Q_mQ_mQ_f$&4$/$3&4$/$3&&1$/$3&0&&4$/$3&2$/$3&&3&2\\
&$Q_mQ_mq_f$&4$/$3&4$/$3&&0&1$/$3&&2$/$3&0&&2&5$/$3\\[5pt]
&&$Q_mQ_mQ_f$&$q_mq_mQ_f$&&$Q_mQ_mQ_f$&$q_mq_mQ_f$&&$Q_mQ_mQ_f$&$q_mq_mQ_f$&&$Q_mQ_mQ_f$&$q_mq_mQ_f$  \\
$Qq\times QQ$&$Q_mQ_mQ_f$&4$/$3&0&&1$/$3&1$/$3&&4$/$3&2$/$3&&3&1\\
&$q_mq_mQ_f$&0&4$/$3&&1$/$3&1$/$3&&2$/$3&0&&1&5$/$3\\[5pt]
&&$q_mq_mQ_f$&$q_mq_mq_f$&&$q_mq_mQ_f$&$q_mq_mq_f$ &&$q_mq_mQ_f$&$q_mq_mq_f$&&$q_mq_mQ_f$&$q_mq_mq_f$ \\
$qq\times Qq$&$q_mq_mQ_f$&4$/$3&4$/$3&&1$/$3&0&&0&2$/$3&&5$/$3&2\\
&$q_mq_mq_f$&4$/$3&4$/$3&&0&1$/$3&&2$/$3&4$/$3&&2&3\\[5pt]
&&$Q_mQ_mq_f$&$q_mq_mq_f$&&$Q_mQ_mq_f$&$q_mq_mq_f$ &&$Q_mQ_mq_f$&$q_mq_mq_f$&&$Q_mQ_mq_f$&$q_mq_mq_f$ \\
$Qq\times qq$&$Q_mQ_mq_f$&4$/$3&0&&1$/$3&1$/$3&&0&2$/$3&&5$/$3&1\\
&$q_mq_mq_f$&0&4$/$3&&1$/$3&1$/$3&&2$/$3&4$/$3&&1&3\\
\hline
\end{tabular*}
\end{sidewaystable}

%s2.2 ###
\subsection{The model}

In QTL mapping different line crosses can be combined together to
increase the parameter inference space via a variance components method
[Xie, Gessler and Xu (\citeyear{XGX1998})]. VC method has been shown to be powerful in
assessing genomic imprinting in human linkage analysis [Hanson et al.
(\citeyear{Hetal2001})]. Recently, Li and Cui (\citeyear{LC2009a}) extended the VC
model to experimental crosses and proposed an iQTL mapping framework
via combining different line crosses for iQTL detection. We extend our
previous work to triploid endosperm tissue considering the unique
genetic components in the endosperm genome.

Suppose total $K$ families are collected which are composed of the four
distinct backcross families. Assume $n_{k}$ individuals are sampled in
the $k$th family. The phenotypic variation of a quantitative trait in
family $k$ (denoted as $y_k$) can be explained by the genotype-specific
cytoplasmic maternal effect (denoted as $\mu_k$), additive QTL effect
(denoted as $a_k$), polygene effect (denoted as $g_k$) and random
residual effect (denoted as $e_k$). To incorporate the parent-of-origin
effect, the additive QTL effect ($a_k$) can be further partitioned into
two separate effects, an effect due to the expression of the maternal
allele (denoted as $a_{km}$) and an effect due to the expression of the
paternal allele (denoted as $a_{kf}$). The model can thus be expressed
as
%e1 ###
\begin{equation}\label{model1}
\qquad y_{ki}=\mu_{k}+2a_{kmi}+a_{kfi}+g_{ki}+e_{ki}, \qquad k=1,\ldots,K;  i=1,\ldots,n_k,
\end{equation}
where ${a}_{kmi}$, ${a}_{kfi}$, ${g}_{ki}$ and ${e}_{ki}$ are random
effects with normal distribution, that is, ${a}_{kmi}\sim
N({0},\pi_{i_mj_m|k}\sigma_{m}^2)$, ${a}_{kfi}\sim
N({0},\pi_{i_m/j_f|k}\sigma_{f}^2)$, ${g}_{ki}\sim
N({0},\phi_{ij|k}\sigma_{g}^2)$, ${e}_{ki}\sim N({0},\sigma_{e}^2)$;
${g}_{ki}$ and ${e}_{ki}$ are uncorrelated to ${a}_{kmi}$ and
${a}_{kfi}$; the coefficient 2 for ${a}_{kmi}$ adjusts for the effects
of two identical maternal copies; $\mu_{k}$ models the maternal
genotype-specific effect; $\pi_{i_mj_m|k}$, $\pi_{i_fj_f|k}$ and
$\phi_{ij|k}$ are the IBD coefficients which are explained in the
following section. With four distinct segregation populations, we have
only three distinct maternal genotypes, $AA$, $Aa$ and $aa$. Thus, the
parameter $\mu_k$ can be collapsed into three distinct values denoted
as $\mu_1$, $\mu_2$ and $\mu_3$ corresponding to maternal genotypes
$AA$, $Aa$ and $aa$, respectively. Letting $\beta=(\mu_1, \mu_2,
\mu_3)$, then model (\ref{model1}) can be rewritten in a vector form as
%e2 ###
\begin{equation}\label{model2}
\mathbf{y}_{k}=X_{k}\beta+2\mathbf{a}_{km}+\mathbf{a}_{kf}+\mathbf{g}_{k}+\mathbf{e}_{k},\qquad  k=1,\ldots,K,
\end{equation}
where $X_k$ is an $n_k\times 3$ matrix with one column of ones and two columns of zeros.

%s2.3 ###
\subsection{Parent-specific allele sharing and the covariance between two inbreeding sibs}\label{sec2.3}
One of the major tasks in IBD-based iQTL mapping with the variance
components model is to calculate the IBD sharing probabilities and the
phenotypic covariances between sibs. Such a method has been developed
in the human population [Hanson et al. (\citeyear{Hetal2001})], which, however, cannot
be applied to a complete inbreeding population in experimental crosses,
because the allelic sharing relationship among sibpairs does not follow
the pattern as the one derived from a natural noninbreeding population.
Instead, the IBD sharing probability can be calculated based on
Mal\'{e}cot's coefficient of coancestry (\citeyear{M1948}) for an inbreeding
population. Li and Cui (\citeyear{LC2009a}) recently explored different
allelic sharing patterns among sibpairs in a reciprocal backcross
design with a diploid tissue. We extend the method to the triploid
endosperm genome and derive covariances among sibpairs in a triploid
tissue.

%f1 ###
\begin{figure}

\includegraphics{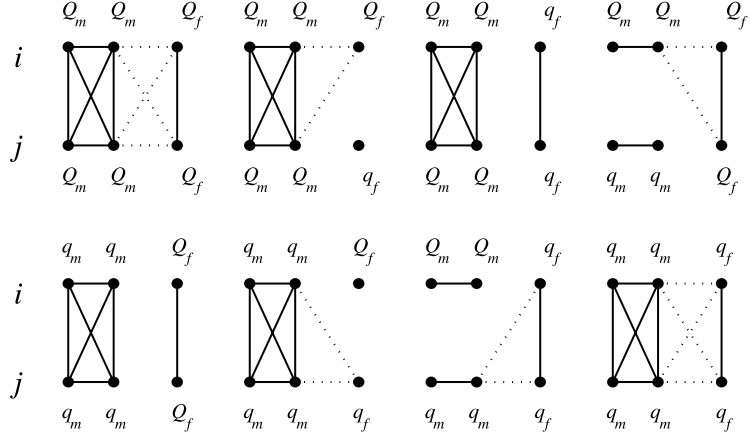}

\caption{Possible alleles shared IBD for individuals $i$ and
$j$ in inbreeding backcross families. The solid lines indicate IBD
sharing for alleles inherited from the same parent. The dotted lines
indicate IBD cross-sharing for alleles inherited from different
parents.}\label{fIBD}
\end{figure}

Consider two individuals $i$ and $j$ randomly selected from one
backcross family with phenotype $y_i$ and $y_j$. Figure \ref{fIBD}
shows all possible allelic sharing patterns between individuals $i$ and
$j$. The solid line indicates IBD sharing for alleles derived from the
same parent and the dotted line indicates IBD cross-sharing for alleles
derived from different parents. The allelic cross-sharing is unique to
inbreeding populations, whereby this cross-sharing probability reduces
to zero for noninbreeding populations. Here we propose to calculate the
IBD sharing between individuals $i$ and $j$ (denoted as $\pi_{ij}$) for
a triploid genome as
%e3 ###
\begin{equation}\label{IBDt}
\pi_{ij}= \cases{
3\theta_{ij},& \quad if $ i\neq j$,\cr
\frac{1}{3}(5+3F_i), &\quad if $i=j$,
}
\end{equation}
where $\theta_{ij}$ is Mal\'{e}cot's coefficient of coancestry and
$F_i$ is the inbreeding coefficient [Harris (\citeyear{H1964}); Cockerham (\citeyear{C1983});
Lynch and Walsh (\citeyear{LW1998})]. By definition, $\theta_{ij}$ is calculated as
the probability of two randomly selected alleles from individuals $i$
and $j$ being identical by descent. The calculation of $\pi_{ij}$ is
different from the usual IBD sharing calculation in noninbreeding
populations. It is instead interpreted as triple the Mal\'{e}cot
coefficient of coancestry [Xie, Gessler and Xu (\citeyear{XGX1998})]. For easy notation, we
still adopt the term ``IBD sharing probability'' for $\pi_{ij}$ in the
rest of the presentation. The calculation of the inbreeding coefficient
follows the procedure given in Lynch and Walsh (\citeyear{LW1998}).

To illustrate the idea, consider two backcross individuals $i$ (with
genotype $A_mA_mA_f$) and $j$ (with genotype $B_mB_mB_f$). The
coefficient of coancestry $\theta_{ij}$ between these two individuals
can be expressed as
\begin{eqnarray*}
\theta_{ij}&=&\tfrac{1}{9}\{\Pr(A_{m1}\equiv B_{m1})+\Pr(A_{m1}\equiv B_{m2})+\Pr(A_{m2}\equiv B_{m1})\\
& &\hphantom{\tfrac{1}{9}\{}{}+\Pr(A_{m2}\equiv B_{m2})+\Pr(A_{m1}\equiv B_f)+\Pr(A_{m2}\equiv B_f)\\
& &\hphantom{\tfrac{1}{9}\{}\hspace*{9pt}{}+\Pr(A_f\equiv B_{m1})+\Pr(A_f\equiv B_{m2})+\Pr(A_f\equiv B_f)\}\\
&=&\tfrac{1}{9}(4\theta_{i_mj_m}+2\theta_{i_mj_f}+2\theta_{i_fj_m}+\theta_{i_fj_f}),
\end{eqnarray*}
where the notation $\equiv$ refers to identical by decent; the
subscript numbers 1 and 2 indicate two maternally inherited alleles;
$\theta_{i\cdot j\cdot}$ is defined as the allelic kinship coefficient [Lynch
and Walsh (\citeyear{LW1998})]. Note that the two terms $\theta_{i_mj_f}$ and
$\theta_{i_fj_m}$ are indistinguishable, but their sum denoted as
$\theta_{i_m/j_f}(=\theta_{i_mj_f}+\theta_{i_fj_m})$ is unique. Thus,
we have
$\theta_{ij}=\frac{1}{9}(4\theta_{i_mj_m}+2\theta_{i_m/j_f}+\theta_{i_fj_f})$.
Following equation (\ref{IBDt}), we have
\[
\pi_{ij}=3\theta_{ij}=\tfrac{4}{3}\theta_{i_mj_m}+\tfrac{2}{3}\theta_{i_m/j_f}+\tfrac{1}{3}\theta_{i_fj_f}=\pi_{i_mj_m}+\pi_{i_m/j_f}+\pi_{i_fj_f}
\qquad\mbox{for }  i\neq j.
\]
It can be seen that the IBD sharing between any two individuals can be
decomposed as three separate components, one due to the IBD sharing for
alleles derived from the maternal parent
($\pi_{i_mj_m}=\frac{4}{3}\theta_{i_mj_m}$), one due to the
cross-sharing for alleles derived from different parents
($\pi_{i_m/j_f}=\frac{2}{3}\theta_{i_m/j_f}$) and one due to the IBD
sharing for alleles derived from the paternal parent
($\pi_{i_fj_f}=\frac{1}{3}\theta_{i_fj_f}$). An exhaustive list of all
possible IBD sharing probabilities for the four backcross families is
given in Table \ref{IBD}.

Dropping the family index $k$, the covariance between any two individuals $i$ and $j$ can be expressed as
\begin{eqnarray*}
&&\operatorname{Cov}(y_i, y_j|\pi_{i_mj_m},\pi_{i_m/j_f},\pi_{i_fj_f})\\
&&\qquad=\operatorname{Cov}(2a_{mi}+a_{fi}+g_i+e_i,  2a_{mj}+a_{fj}+g_j+e_j)\\
&&\qquad=4\pi'_{i_mj_m}\sigma^2_m+2\pi'_{i_m/j_f}\sigma^2_{mf}+\pi_{i_fj_f}\sigma^2_f+\phi_{ij}\sigma^2_g+I_{ij}\sigma^2_e,
\end{eqnarray*}
where $\pi'_{i_mj_m}=\frac{1}{4}(\pi_{i_mj_m})$ and
$\pi'_{i_m/j_f}=\frac{1}{2}(\pi_{i_m/j_f})$ are the IBD sharing and
cross-sharing probabilities by considering one single maternal allele;
$\sigma^2_{mf}$ measures the variation of trait distribution due to
alleles cross-sharing; $\phi_{ij}$ is the expected alleles shared IBD;
$I_{ij}$ is an indicator variable taking value 1 if $i=j$ and 0 if
$i\neq j$. For a natural population without inbreeding, there is no
allele cross-sharing for an individual with itself, hence,
$\pi_{i_m/j_f}=0$. For a diploid noninbreeding population, the trait
covariance can be simplified as the one given in Shete, Zhou and Amos
(\citeyear{SZA2003}). In matrix form, the phenotypic
variance-covariance for individuals in the $k$th backcross family can
then be expressed as
%e4 ###
\begin{equation}\label{vci}
\bolds{\Sigma}_k=\bolds{\Pi}_{m|k}\sigma_{m}^2+\bolds{\Pi}_{m/f|k}\sigma_{mf}^2+\bolds{\Pi}_{f|k}\sigma_{f}^2+\bolds{\Phi}_{g|k}\sigma_g^2+\mathbf{I}\sigma_e^2,
\end{equation}
where the elements of $\bolds{\Pi}_{m|k}$, $\bolds{\Pi}_{f|k}$ and $\bolds{\Pi}_{m/f|k}$ can be found in Table \ref{IBD}.

%s2.4 ###
\subsection{QTL IBD sharing and genome-wide linkage scan}
The above described IBD sharing probability is calculated at a known
marker position. Unless markers are dense enough, we have to search
across the genome for potential (i)QTL positions and their effects. In
general, the QTL position can be viewed as a fixed parameter by
searching for a putative QTL at every 1 or 2 cM on a map interval
bracketed by two markers throughout the entire linkage map. Thus, we
need to estimate the QTL IBD sharing at every scan position. Since the
conditional probability of an endosperm QTL given upon two flanking
markers is the same as the one derived from a diploid genome [Cui and
Wu (\citeyear{CW2005})], the same procedure termed as the expected conditional IBD
sharing described in Li and Cui (\citeyear{LC2009a}) can be applied to
calculate the QTL IBD sharing probability at every scan position.

Assuming multivariate normality of the trait distribution for data in
each family and assuming independence between families, the joint
log-likelihood function when $K$ backcross families are sampled can be
formulated as
%e5 ###
\begin{equation}\label{loglike1}
\ell=\sum_{k=1}^K \log [f(\mathbf{y}_k; {\mu_k}, \bolds{\Sigma}_k)],
\end{equation}
where $f$ is the multivariate normal density. Parameters to be
estimated include ${\beta}=(\mu_1, \mu_2, u_3)$ and
$\Omega=(\sigma_m^2, \sigma_f^2, \sigma_{mf}^2, \sigma_g^2,
\sigma_e^2)$. Two commonly used methods in linkage analysis, the
maximum likelihood (ML) method and the restricted maximum likelihood
(REML) method, may be applied to estimate parameters. It is commonly
recognized that the REML method gives less biased estimation compared
to the ML method [Corbeil and Searle (\citeyear{CS1976})]. Here we adopt the REML
method with the Fisher scoring algorithm to obtain the REML estimates
[see Li and Cui (\citeyear{LC2009a}) for details of the algorithm].

The conditional QTL IBD-sharing values vary at different testing
positions. The amount of support for a QTL at a particular map position
can be displayed graphically through the use of likelihood ratio
profiles, which reflect the variation of the testing position of
putative QTLs. The significant QTLs are detected by the peaks of the
profile plot that pass a certain significant threshold (see Section \ref{sec2.5}
for more details).

%s2.5 ###
\subsection{Hypothesis testing}\label{sec2.5}
In iQTL mapping, we are interested in testing whether there is any
significant genetic effect at a test position and would like to further
quantify the imprinting effect if any. The hypothesis for testing the
existence of a QTL can be expressed as
%e6 ###
\begin{equation}\label{OT}
 \cases{
 H_0\dvtx \sigma_m^2=\sigma_f^2=\sigma^2_{mf}=0,\cr
H_1\dvtx \mbox{at least one parameter is not zero.}
}
\end{equation}
The LRT is applied for this purpose. Define $\widetilde{\Omega}$ and
$\widehat{\Omega}$ to be the estimates of the unknown parameters under
$H_0$ and $H_1$, respectively. The LRT statistic can be calculated as
%e7 ###
\begin{equation}\label{LRT}
\mathrm{LR}=-2[\log L(\widetilde{\Omega}|\mathbf{y})-\log L(\widehat{\Omega}|\mathbf{y})].
\end{equation}

Let $\bolds{\theta}=({\mu_1 \enskip\! \mu_2 \enskip\! \mu_3\enskip\! \theta_1\enskip\! \theta_2 \enskip\! \theta_3
\enskip\!\theta_4 \enskip\! \theta_5})^T=({\mu_1 \enskip\! \mu_2  \enskip\!\mu_3 \enskip\!\sigma_m^2  \enskip\!\sigma_f^2\enskip\!
\sigma_{mf}^2  \enskip\!\sigma_g^2 \enskip\!\sigma_e^2})^T\in
{\Omega}=\mathbb{R}^3\times[0,\infty)\times[0,\infty)\times[0,\infty)\times(0,\infty)\times(0,\infty)$
be the parameters to be estimated. Note that the polygene variance is
bounded away from zero if we assume there are more than one QTL in the
genome. Let the true parameters under the null hypothesis be
$\bolds{\theta}_0=(\mu_{10} \enskip\mu_{20} \enskip\mu_{30} \enskip\sigma^2_{m_0}\enskip
\sigma^2_{f_0} \enskip\sigma^2_{mf_0} \enskip \sigma^2_{g_0}\enskip
\sigma^2_{e_0})^T=(\mu_{10} \enskip\mu_{20}\enskip \mu_{30} \enskip0\enskip 0\enskip 0\enskip  \sigma^2_{g_0}\enskip
\sigma^2_{e_0})^T\in\Omega_0=\mathbb{R}^3\times\{0\}\times\{0\}\times\{0\}\times(0,\infty)\times(0,\infty)$.
The three tested genetic variance components under the null hypothesis
lie on the boundaries of the parameter space ${\Omega}$. Thus, the
standard conditions for obtaining the asymptotic $\chi^2$ distribution
of the LRT are not satisfied [Self and Liang (\citeyear{SL1987})]. Following the
results from Chernoff (\citeyear{C1954}), Shapiro (\citeyear{S1985}) and Self and Liang (\citeyear{SL1987}),
the following theorem states that the LR statistic follows a mixture
chi-square distribution, whereby the mixture proportions depend on the
estimated Fisher information matrix.

\begin{thm}\label{thm1}
Let $C_{\Omega_0}$ and $C_{\Omega}$ be closed convex cones with vertex
at $\bolds{\theta}_0$ to approximate $\Omega_0$ and $\Omega$,
respectively. Let $\mathbf{Y}$ be a random variable with a multivariate
normal distribution with mean $\bolds{\theta}_0$, and
variance--covariance matrix $I^{-1}(\bolds{\theta}_0)$. Under the
assumptions given in the \hyperref[app]{Appendix}, the LR statistic in (\ref{LRT}) is
asymptotically distributed as a mixture chi-square distribution with
the form
$\omega_3\chi^2_3\dvtx\omega_2\chi^2_2\dvtx\omega_1\chi^2_1\dvtx\omega_0\chi^2_0$,
where
$\omega_3=\frac{1}{4\pi}[2\pi-\cos^{-1}\rho_{12}-\cos^{-1}\rho_{13}-\cos^{-1}\rho_{23}]$,
$\omega_2=\frac{1}{4\pi}[3\pi-\cos^{-1}\rho_{12|3}-\cos^{-1}\rho_{13|2}-\cos^{-1}\rho_{23|1}]$,
$\omega_1=\frac{1}{4\pi}(\cos^{-1}\rho_{12}+\cos^{-1}\rho_{13}+\cos^{-1}\rho_{23})$,
and
$\omega_0=\frac{1}{2}-\frac{1}{4\pi}[3\pi-\cos^{-1}\rho_{12|3}-\cos^{-1}\rho_{13|2}-\cos^{-1}\rho_{23|1}]$;
$\rho_{ab}$ is the correlation between the variance terms $a$ and $b$
calculated from the Fisher information matrix, and
$\rho_{ab|c}=\break\frac{(\rho_{ab}-\rho_{ac}\rho_{bc})}{(1-\rho_{ac}^2)^{1/2}(1-\rho_{bc}^2)^{1/2}}$.
\end{thm}

Note that the symbol $\pi$ in the above theorem is the irrational
number (a mathematical constant) not the IBD sharing probability. The
proof of the theorem is given in the \hyperref[app]{Appendix}.

\begin{remark*}
When the random parameter estimators are uncorrelated or the
correlation is extremely small, that is, the Fisher information matrix
is close to diagonal, the mixture proportions for the $\chi_k^2$
components are reduced to the binomial form with ${3\choose k}2^{-3}$, which is consistent with the result
(Case 9) given in Self and Liang (\citeyear{SL1987}).
\end{remark*}

Once a QTL is identified at a genomic position, we can further assess
its imprinting property. To evaluate whether a QTL shows imprinting
effect, the hypotheses can be formulated as
%e8 ###
\begin{equation}\label{iT}
 \cases{
 H_0\dvtx \sigma^2_f=\sigma^2_m,\cr
H_1\dvtx \sigma^2_f \neq \sigma^2_m.
}
\end{equation}
Again, the likelihood ratio test can be applied which asymptotically
follows a $\chi^2$ distribution with 1 degree of freedom since the
tested parameter under the null is nonnegative and does not lie on the
boundary of the parameter space. Rejecting $H_0$ indicates genomic
imprinting, and the QTL can be called an iQTL. We denote this
imprinting test as $\mathrm{LR}_{\mathrm{imp}}$. If the null is
rejected, one would be interested in testing whether the detected iQTL
is completely maternally or paternally imprinted with the corresponding
null hypothesis expressed as $H_0\dvtx \sigma^2_m=0$ and $H_0\dvtx
\sigma^2_f=0$, respectively. The LRT statistic for the two tests
asymptotically follows a mixture $\chi^2$ distribution with the form
$\frac{1}{2}\chi^2_0\dvtx\frac{1}{2}\chi^2_1$. Rejection of complete
imprinting indicates partial imprinting.

Maternal effects can be tested by formulating hypothesis: $H_0:
\mu_1=\mu_2=\mu_3$. Note that these three parameters do not represent
the true maternal effects, as they are confounded with the main genetic
effects. But a test of pairwise differences can be applied to detect
the significance of any maternal contribution.\looseness=1

%s2.6 ###
\subsection{Multiple iQTL model}
In practice, there may be several QTLs to reflect the phenotypic
variation in the whole genome. When testing QTL effects at one
chromosome, effects from QTLs located at other chromosomes are absorbed
by the polygenic effect ($g$). In some cases, two or more QTLs may be
located at the same chromosome, which are termed as background QTL(s)
in comparison to the tested one. When this happens, it is essential to
adjust for the background QTL(s)' effects. Otherwise, it may lead to
biased estimation for the putative QTL caused by the interference of
QTL(s) close to the tested interval [Zeng (\citeyear{Z1994})].

In the previous work of Li and Cui (\citeyear{LC2009a}), the authors
proposed a multiple iQTL model following the idea of next-to-flanking
markers proposed by Xu and Atchley (\citeyear{XA1995}). We adopted a similar
strategy in the current study. Briefly, assuming there are $S$ (i)QTLs
in one chromosome, the multiple iQTL model considering parent-specific
allele effect can be expressed as
\[\label{mqtl1}
y_{ki}=\mu_k+\sum_{s=1}^S2a_{kmis}+\sum_{s=1}^Sa_{kfis}+g_{ki}+e_{ki}, \qquad k=1,\ldots,K;  i=1,\ldots,n_k,
\]
where each (i)QTL effect is partitioned as two separate terms to
reflect the contribution of the maternal and paternal alleles. In
reality, the exact number and location of QTLs in a chromosome is
generally unknown before doing a genome-wide search. This problem can
be eased by applying the next-to-flanking markers idea proposed by Xu and Atchley (\citeyear{XA1995}).

Denote a test interval with two flanking markers as
$\mathcal{M}_l$--$\mathcal{M}_{r}$. The markers next to these two
markers are denoted as $\mathcal{M}_{L}$ on the left of
$\mathcal{M}_l$, and $\mathcal{M}_{R}$ on the right of
$\mathcal{M}_{r}$ ($L=l-1$ and $R=r+1$). Conditional on the two
markers, $\mathcal{M}_{L}$ and $\mathcal{M}_{R}$, we expect the effects
of QTL(s) located outside of the tested interval can be absorbed by the
IBD values calculated from the two next-to-flanking markers [Xu and
Atchley (\citeyear{XA1995})]. Thus, the calculation of (i)QTL
covariance conditional on these two markers will avoid the requirement
for the position of QTLs outside of the tested interval. Dropping the
family index, the phenotypic covariance between two individuals $i$ and
$j$ can be expressed as
\begin{eqnarray*}
&&\operatorname{Cov}(y_{i},y_{j}|\pi_{{L}},\hat{\pi}_{i_mj_m},\hat{\pi}_{i_m/j_f}, \hat{\pi}_{i_fj_f},\pi_{R})\\
&&\quad=\sum_{l=1}^{L}K(\theta_{l{L}},\pi_{{L}})\sigma_{l}^2+\hat{\pi}_{i_mj_m}\sigma_m^2+\hat{\pi}_{i_m/j_f}\sigma_{mf}^2+\hat{\pi}_{i_fj_f}\sigma_f^2\\
&&\qquad{}+\sum_{r=1}^{R}K(\theta_{l{R}},\pi_{{R}})\sigma_{r}^2+\phi_{ij}\sigma^2_g+I_{ij}\sigma^2_e \\
&&\quad=\pi_{{L}}\sigma_L^2+\hat{\pi}_{i_mj_m}\sigma_m^2+\hat{\pi}_{i_m/j_f}\sigma_{mf}^2+\hat{\pi}_{i_fi_f}\sigma_f^2+\pi_{{R}}\sigma_R^2+\phi_{ij}\sigma^2_g+I_{ij}\sigma^2_e,
\end{eqnarray*}
where $\pi_{{L}}$ is the IBD sharing value at marker $L$, and
$\sigma_L^2$ is a composite variance component which reflects the
variation of (i)QTL effects on the left side of the tested interval
[see Li and Cui (\citeyear{LC2009a}) for details]. $\pi_{{R}}$ and
$\sigma_R^2$ are defined similarly. The calculations of $\pi_{{L}}$ and
$\pi_{{R}}$ reflect the triploid structure of the endosperm genome.
Testing (i)QTL effects can then be focused on a tested interval while
adjusting for the background QTLs' effects located in another place.

%s3 ###
\section{Simulation}\label{sec3}
Simulation studies are conducted to investigate the\break method performance.
We assume a fixed total sample size of 400, then vary the family and
offspring size with different combinations, that is, $4\times100$,
$8\times50$, $20\times20$ and $100\times4$, in order to evaluate the
effect of family and offspring size on testing power and parameter
estimation. Simulation details are given in the \ref{suppA}. Here we
briefly summarize the main results.

%s3.1 ###
\subsection{Single iQTL simulation}
For the single iQTL simulation, the results show that both the
$4\times100$ and the $100\times4$ designs yield lower QTL detection
power and higher RMSE (root mean squared error) for QTL position
estimation than the other two designs do. The $20\times20$ design
slightly beats the $8\times50$ design with smaller imprinting type I
error and higher QTL detection power. These results indicate that it is
necessary to maintain a balance between the family size and the
offspring size, in order to achieve optimal power and good effects
estimation precision. For a given budget with a fixed total sample
size, one should always try to avoid extreme designs with a large (or
small) number of families, each with a small (or large) number of
offsprings.

Focusing on the $20\times20$ design, simulations are performed to show
the model behavior under different imprinting modes, that is, complete
paternal imprinting, complete maternal imprinting, partial maternal
imprinting and partial paternal imprinting. The results indicate that
the power to detect imprinting depends on the underlying degree of
imprinting. Relatively higher imprinting power is observed when an iQTL
is maternally imprinting compared to the case when an iQTL is
paternally imprinting.

%s3.2 ###
\subsection{Multiple iQTL simulation}
In this simulation data are simulated by assuming two (i)QTLs located
at two genomic positions and are subject to both the single iQTL and
multiple iQTL analyses. The results indicate a clear benefit of
analysis by fitting a multiple iQTL model rather than fitting a single
iQTL model. While the single iQTL analysis detects one ``ghost'' QTL
located between the two simulated QTLs, the multiple iQTL analysis can
clearly separate the two QTLs with high precision. Note that the
multiple iQTL analysis normally generates lower LR values than the
single iQTL analysis does. Note that the distribution of the LR value
under the multiple iQTL analysis is not clear, and permutation should
be applied to assess significance of any (i)QTLs in multiple iQTL
analysis [Xu and Atchley (\citeyear{XA1995})].

%s4 ###
\section{A case study}\label{sec4}
We apply our method to a real data set which has two endosperm traits
of interests: mean ploidy level (denoted as Mploidy) and percentage of
endoreduplicated nuclei (denoted as Endo). The two traits describe the
level of endoreduplication in maize endosperm, which is thought to be
genetically controlled by imprinted genes [Dilkes et al. (\citeyear{Detal2002})]. Four
backcross (BC) segregation populations, initiated with two inbred
lines, Sg18 and Mo17, were sampled. The four BC populations were
obtained following the design illustrated in Table \ref{IBD}. The data
show a large degree of variation for endoreduplication among the four
BC populations, and ten linkage groups were constructed from the
observed marker data [Coelho et al. (\citeyear{Cetal2007})]. Readers are referred to
Coelho et al. (\citeyear{Cetal2007}) for more details about the data. The two traits
are analyzed with our multiple iQTL model aimed to identify iQTLs
across the ten linkage groups. The data are also analyzed with a
Mendelian model. Results from both imprinting and Mendelian models are
compared and summarized in the Supplementary Materials. %\ref{suppA}.

%f2 ###
\begin{figure}

\includegraphics{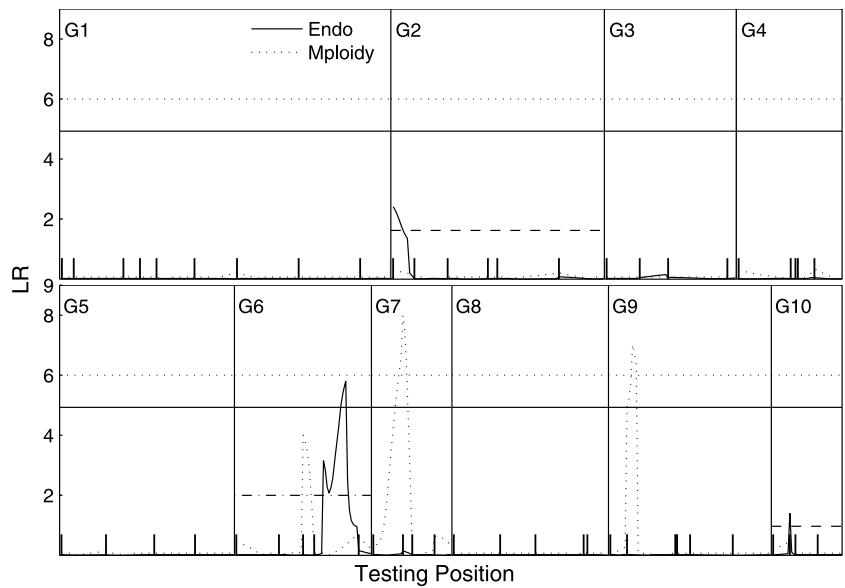}

\caption{The profile of the log-likelihood ratios (LR) for testing the
existence of QTLs underlying the two endosperm traits across the 10
maize linkage groups ($\mathrm{G}_1,\ldots,\mathrm{G}_{10}$). The
genome-wide LR profiles for the percentage of endoreduplication (Endo)
and mean ploidy (Mploidy) traits are indicated by solid and dotted
curves, respectively. The threshold values for claiming the existence
of QTLs are given as the horizonal solid and dotted line for the
genome-wide threshold,  and the dashed and dash-dotted line for the
chromosome-wide threshold, for the two traits Endo and Mploidy,
respectively. The genomic positions corresponding to the peak of the
curves that pass the corresponding thresholds are the MLEs of the QTL
location. The positions of markers on the linkage groups [Coelho et al.
(\protect\citeyear{Cetal2007})] are indicated at ticks.}\label{LR}
\end{figure}

Figure \ref{LR} plots the LR values across the ten linkage groups for
the two traits. The solid and dotted curves represent LR profiles for
traits Endo and Mploidy, respectively. To adjust for the genome-wide
error rate across the entire linkage group, permutation tests are
applied in which the critical threshold value is empirically calculated
on the basis of repeatedly shuffling the relationships between marker
genotypes and phenotypes within each BC family [Churchill and Doerge
(\citeyear{CD1994})]. The corresponding genome-wide significance thresholds (at 5\%
level) for the two traits are denoted by the horizontal solid (for
Endo) and dotted (for Mploidy) lines. The 5\% level chromosome-wide
thresholds are denoted by the dashed (for Endo) and dash-dotted (for
Mploidy) lines. QTLs that are significant at the chromosome-wide level
are called suggestive QTLs. It can be seen that two QTLs (on G7 and G9)
associated with Mploidy and one QTL (on G6) associated with Endo are
detected at the 5\% genome-wide significance level (denoted by ``$*$''
in Table \ref{real}). Two suggestive QTLs (on G2 and G10) associated
with Endo and one suggestive QTL (on G6) associated with Mploidy are
also identified. The detailed QTL location and effect estimates as well
as the test results for imprinting are tabulated in Table~\ref{real}.
For the trait Mploidy, the identified three QTLs are all imprinted
($p_{\mathrm{imp}}<0.05$) and all show completely maternal imprinting, that is,
the maternal copy does not express. They are thus termed iQTLs. The
cytoplasmic maternal effect does not show any evidence of significance
for all the three iQTLs ($p_M>0.05$). For the trait Endo, only the QTL
detected on G6 shows imprinting effect ($p_{\mathrm{imp}}<0.05$) and it shows
completely paternal imprinting ($p_{f}<0.05$). The other two QTLs do
not show evidence of imprinting ($p_{\mathrm{imp}}>0.05$). For this trait,
significant maternal effects are detected ($p_M<0.01$).

%t2 ###
\begin{sidewaystable}
\tablewidth=\textheight
\tablewidth=\textwidth
\caption{The estimated parameters for the three maternal effects and
the variance components for two endosperm traits: mean ploidy (Mploidy)
and percent of the endoreduplicated nuclei (Endo)}\label{real}
\begin{tabular*}{\textwidth}{@{\extracolsep{\fill}}lcd{2.2}d{2.2}d{2.2}d{2.2}d{2.2}d{1.2}d{2.2}d{2.2}d{1.2}d{2.2}d{2.2}d{1.3}d{1.3}d{1.2}@{}}
\hline
&&\multicolumn{3}{c}{\textbf{Maternal effects}}& \multicolumn{3}{c}{\textbf{Genetic effects}}&&&&&&\\[-5pt]
&&\multicolumn{3}{c}{\hrulefill}& \multicolumn{3}{c}{\hrulefill}\\
\textbf{Trait} &\textbf{Ch}&\multicolumn{1}{c}{$\bolds{\mu_1}$}&\multicolumn{1}{c}{$\bolds{\mu_2}$}&\multicolumn{1}{c}{$\bolds{\mu_3}$}&\multicolumn{1}{c}{$\bolds{\sigma^2_{m}}$}&
\multicolumn{1}{c}{$\bolds{\sigma^2_{f}}$}&\multicolumn{1}{c}{$\bolds{\sigma^2_{mf}}$} &\multicolumn{1}{c}{$\bolds{\sigma^2_L}$}&\multicolumn{1}{c}{$\bolds{\sigma^2_R}$}&
\multicolumn{1}{c}{$\bolds{\sigma^2_{g}}$}&\multicolumn{1}{c}{$\bolds{\sigma^2_{e}}$}&\multicolumn{1}{c}{$\bolds{p_M}$}&\multicolumn{1}{c}{$\bolds{p}_{\mathbf{imp}}$}&
\multicolumn{1}{c}{$\bolds{p_{m}}$}&\multicolumn{1}{c@{}}{$\bolds{p_{f}}$}\\
\hline
Mploidy&\hphantom{0}6$^*$&13.13&11.88&9.78&0.01&0.30&0.03& {\approx}0 &0.22&1.25&2.59&0.34&0.045&0.023&0.31\\
           &7&11.78&11.19&9.16&0.15&0.60&0.94& {\approx}0 &0.12&1.07&2.69&0.31&0.048&0.024&0.49\\
           &9&13.84&12.08&10.01& {\approx}0 &0.94&0.71& {\approx}0 &0.01&1.59&2.55&0.12&0.013&0.021&0.48\\[5pt]
Endo&\hphantom{0}2$^*$&72.23&62.40&52.86&0.43&0.83&2.41&0.99& {\approx}0 &5.10&37.49& {<} 0.01&0.67&\multicolumn{1}{c}{--}&\multicolumn{1}{c@{}}{--}\\
              &6&68.37&63.18&54.92&2.92& {\approx}0 &7.14&1.42&0.92&1.28&38.91& {<} 0.01&0.02&0.28&0.01\\
              &10$^*$&70.78&62.28&50.67&0.58&0.03&1.52& {\approx}0 &0.17&3.24&39.20& {<} 0.01&0.29&\multicolumn{1}{c}{--}&\multicolumn{1}{c@{}}{--}\\
\hline
\end{tabular*}
\legend{The three QTLs for trait Mploidy are located at marker umc1805,
marker dupssr9 and $\mathrm{umc}1040+5.76\mathrm{cM}$ on chromosome 6, 7 and 9,
respectively. The three QTLs for trait Endo are located at marker
umc2094, $\mathrm{bnlg}345+33.49\mathrm{cM}$ and $\mathrm{MMC}501+18\mathrm{cM}$ on chromosome 2, 6 and 10,
respectively. QTLs showing significance at the genome-wide significance
level are indicated by ``$_*$''. $p_M$, $p_{\mathrm{imp}}$, $p_{m}$ and
$p_{f}$ are the $p$-values for testing maternal effect
($H_0\dvtx\mu_1=\mu_2=\mu_3$), imprinting effect ($H_0\dvtx
\sigma_{m}^2=\sigma_f^2$), complete maternal imprinting ($H_0\dvtx
\sigma_m^2=0$) and complete paternal ($H_0\dvtx \sigma_f^2=0$),
respectively.}
\end{sidewaystable}

In our study, one maternally controlled iQTL is detected for trait
Endo, which is consistent with the result given by Dilkes et al.
(\citeyear{Detal2002}). Meanwhile, according to the genetic conflict theory proposed by
Haig and Westoby (\citeyear{HW1991}), maternally derived alleles tend to trigger a
negative effect on the increase of endosperm growth, whereas paternally
derived alleles tend to play an opposite effect to increase seed size.
The identified iQTLs showing maternal imprinting for trait Mploidy can
be well explained by the genetic conflict theory. Both empirical
evidence and theoretical hypothesis support the current finding.

%s5 ###
\section{Discussion}\label{sec5}
The role of genomic imprinting in endosperm development has been
commonly recognized [Dilkes et al. (\citeyear{Detal2002}); Kinoshita et al. (\citeyear{Ketal1999});
Chaudhuri and Messing (\citeyear{CM1994})]. But little is known about the exact
location and effect size of imprinted genes in endosperm. As endosperm
in cereal provides the most nutrition for human beings, it is important
to identify imprinted genes that govern seed development, particularly
endosperm development. In this article we develop a variance components
linkage analysis method with an experimental cross design, aimed to
identify iQTLs in endosperm. Our method is motivated by real
applications and is evaluated through Monte Carlo simulations.

The proposed method is based on a particular genetic design (reciprocal
BC design) with inbreeding populations. We treat iQTL effects as
random, different from a fixed-effect iQTL model [e.g., Cui (\citeyear{C2007})].
Variance components linkage analysis with a partial inbreeding human
population was previously proposed [see Abney,  McPeek and Ober (\citeyear{ASO2000})]. However,
extending the VC model to a completely inbreeding population is
challenging. In our previous work, we proposed a VC-based iQTL mapping
framework for an inbreeding diploid mapping population [Li and Cui
(\citeyear{LC2009a})]. Extending the previous work, we propose a novel
IBD partitioning approach to calculate allelic sharing in an inbreeding
endosperm population. Extension to mapping multiple iQTLs is provided.
Simulations indicate good performance of the multiple iQTL analysis
compared to a single iQTL model. Meanwhile, to obtain a good balance of
iQTL position and effect estimation as well as detection power, we have
to avoid extreme sample designs. For a fixed total sample size,
extremely large or small families should be always avoided.

In an application to two endosperm traits, we identified three iQTLs
for trait Mploidy. All show paternal expression. We also identified one
iQTL for trait Endo, which shows a maternal expression. According to
the parental conflict theory proposed by Haig and Westoby (\citeyear{HW1991}),
maternally derived alleles trigger a negative effect on endosperm cell
growth and inhibit endosperm development because the extra maternal
copy could slower nuclear division in endosperm. On the contrary,
paternally derived alleles tend to increase seed size. Thus, the three
iQTLs identified for Mploidy can be explained by the genetic conflict
theory. The occurrence of parental conflict theory explains
parent-of-origin effects as an ubiquitous mechanism for the control of
early seed development [Grossniklaus et al. (\citeyear{Getal2001}); Kinoshita et al.
(\citeyear{Ketal1999})].

In VC-based linkage analyses, likelihood ratio test (LRT) has been
commonly applied in assessing QTL significance. The LRT statistic
asymptotically follows a mixture $\chi^2$ distribution with binomial
mixture coefficients, as many investigators often claimed [following
Case 9 in Self and Liang (\citeyear{SL1987})]. In a recent
investigation, we found that the LRT in a regular VC-based linkage
analysis without considering imprinting follows a mixture $\chi^2$
distribution with mixture proportions depending on the estimated Fisher
information matrix [Li and Cui (\citeyear{LC2009b})]. The modified calculation of
mixture proportion does give more reasonable type I error rate than the
one with binomial coefficients. When imprinting is considered, we show
that the limiting distribution of the LRT also follows a mixture
$\chi^2$ distribution, and we adopt the new criterion for power
evaluation. Simulations show that the new criterion gives type I error
closer to the nominal level than the one using binomial coefficients,
and also produces power as good as the later one (data not shown). We
recommend investigators adopt the new criterion in their analysis.

Increasing evidence has suggested that for correlated traits,
multivariate approaches can increase the power and precision to
identify genetic effects in genetic linkage analyses [e.g., Boomsma and
Dolan (\citeyear{BD1998}); Amos and Andrade (\citeyear{AA2001}); Evans (\citeyear{E2002})]. Also, the joint
analysis of multivariate traits can provide a platform for testing a
number of biologically interesting hypotheses, such as testing
pleiotropic effects of QTL and testing pleiotropic vs close linkage.
Moreover, if the putative QTL has pleiotropic effects on several
traits, the joint analysis may perform better than mapping each trait
separately [Jiang and Zeng (\citeyear{JZ1995})]. Multivariate traits appear
frequently in genetic mapping studies. For example, the two endosperm
traits evaluated in this study are highly correlated [Coelho et al.
(\citeyear{Cetal2007})]. We expect joint analysis may provide high mapping resolution
and power for iQTL detection. This will be explored in our future
investigation. A computer code written in R for implementing the
current analysis is available upon request.

\begin{appendix}

\renewcommand{\theequation}{A\arabic{equation}}
\setcounter{equation}{0}

\section*{Appendix}\label{app}
In standard human linkage analysis with a variance components model,
many authors declare that the likelihood ratio statistic follows a
mixture $\chi^2$ distribution with binomial coefficient for each
mixture component [e.g., Amos and Andrade (\citeyear{AA2001}); Hanson et al.
(\citeyear{Hetal2001}); Shete, Zhou and Amos (\citeyear{SZA2003})]. Following Chernoff (\citeyear{C1954}), Shapiro
(\citeyear{S1985}) and Self and Liang (\citeyear{SL1987}), in the following we show that
the mixture proportion actually depends on the estimated Fisher
information matrix.

For a random sample $\mathbf{X}$ with density function $f(\mathbf{x};
\bolds{\theta})$, following Chernoff (\citeyear{C1954}) and Self and Liang (\citeyear{SL1987}), assume that:
\begin{longlist}[(iii)]
\item [(i)] For any true parameter $\bolds{\theta}_{0}$, the
neighborhood of $\bolds{\theta}_{0}$ is closed and the intersection
between this closure and ${\Omega}$ defined in the main text is
also a closed set.
\item [(ii)] The first three derivatives of
$\sum_{i} \log f(x_i;\bolds{\theta})$ with respect to
$\bolds{\theta}$ on the intersection of the neighborhood of $\bolds{\theta}_{0}$ and ${\Omega}$ almost surely exist. Moreover,
$|\frac{\partial^3 \sum \log f}{\partial \theta_i
\partial \theta_j
\partial \theta_k}|< W(\mathbf{x})$ for all $\theta$ on the
intersection, and $E[W(\mathbf{x})] < \infty$.

\item[(iii)] The information matrix
$\mathcal{I}(\bolds{\theta})$ is positive definite on neighborhoods
of $\bolds{\theta}_0$.

\item[(vi)] The set ${\Omega}$ is convex.
\end{longlist}

Assuming the above assumptions, the consistency, weak convergence and
asymptotic normality of the estimators can be established [see Chernoff
(\citeyear{C1954}); Self and Liang (\citeyear{SL1987}); Shapiro
(\citeyear{S1985})]. Here we cite the main results from Chernoff (\citeyear{C1954}), Shapiro
(\citeyear{S1985}) and Self and Liang (\citeyear{SL1987}) to show the asymptotic
distribution of the LRT in our case.

%{\bf Why do we need this?} Let $\hat\bolds{\theta}=h(\mathbf{X})$ be the estimator that maximize $\underset{i=1}{\overset{n}{\sum}}
%and the $\sqrt{n}$ consistency can be derived from Lemma 1 of Chernoff (1954). The weak convergence of the estimators can also
%be obtained (see Chernoff 1954; Self and Liang 1987; Li and Cui 2009b). %Therefore we have $n^{\frac{1}{2}}(h(\mathbf{X})-\bolds{\theta}_{0})$ $\sim$ N(${\bm 0}$, $I^{-1}(\bolds{\theta}_{0})$).

Defining two closed polyhedral convex cones $C_{\Omega_0}$ and
$C_{\Omega_1}$ to approximate $\Omega_0$ and $\Omega_1$ at $\bolds{\theta}_0$, the parameter space under the null hypothesis is
approximated as $C_{\Omega_0}=\{\bolds{\theta}\dvtx\bolds{\theta}\in
\mathbb{R}^3\times\{0\}\times\{0\}\times\{0\}\times(0,\infty)\times(0,\infty)$\},
against $C_{\Omega_1}=\{\bolds{\theta}\dvtx\bolds{\theta}\in
\mathbb{R}^3\times[0,\infty)\times[0,\infty)\times[0,\infty)\times(0,\infty)\times(0,\infty)$\}
under the alternative. Let $\mathbf{Y}'$ be a random variable generated from
the multivariate normal distribution, that is, $\mathbf{Y}'\sim N(\bolds{\theta}_0$, $I^{-1}(\bolds{\theta}_0))$. Following Chernoff [(\citeyear{C1954}),
Theorem 1], the asymptotic distribution of the LRT in (\ref{LRT})
is equivalent to the following quadratic approximation:
%e9 ###
\begin{equation}\label{A1}
\qquad LR^* = \inf_{\bolds{\theta} \in C_{\Omega_0}}(\mathbf{Y}'-\bolds{\theta})'I(\bolds{\theta}_0)(\mathbf{Y}'-\bolds{\theta})-
\inf_{\bolds{\theta} \in C_{\Omega_1}}(\mathbf{Y}'-\bolds{\theta})'I(\bolds{\theta}_0)(\mathbf{Y}'-\bolds{\theta}).
\end{equation}

Subtracting $\bolds{\theta}_0$ from $\mathbf{Y}'$ and $\bolds{\theta}$, the expression in (\ref{A1}) is given by
%e10 ###
\begin{equation}\label{A2}
\qquad LR^* =\inf_{\bolds{\theta} \in C_{\Omega_0}-\bolds{\theta}_{0}}(\mathbf{Y}-\bolds{\theta})'I(\bolds{\theta}_0)(\mathbf{Y}-\bolds{\theta})-
\inf_{\bolds{\theta} \in C_{\Omega_1}-\bolds{\theta}_{0}}(\mathbf{Y}-\bolds{\theta})'I(\bolds{\theta}_0)(\mathbf{Y}-\bolds{\theta}),
\end{equation}
where $\mathbf{Y}=\mathbf{Y}'-\bolds{\theta}_0 \sim N(\mathbf{0}, I^{-1}(\bolds{\theta}_0))$ under the linear transformation.

Let $C^\ddag=(C_{\Omega_1}-\bolds{\theta}_{0})\cap
(C_{\Omega_0}-\bolds{\theta}_{0})^c=\{\bolds{\theta}\dvtx\theta_1>0,\theta_2>0,\theta_3>0\}$,
which is a closed polyhedral convex cone with 3 dimensions.  By the
Pythagoras theorem, the statistic in (\ref{A2}) can be expressed as
%e11 ###
\begin{eqnarray}\label{A3}
LR^* = \inf_{\bolds{\theta} \in C^\ddag}(\mathbf{Y}-\bolds{\theta})'I(\bolds{\theta}_0)(\mathbf{Y}-\bolds{\theta}).
\end{eqnarray}
Let $\mathcal{F}(C^\ddag)$ be the set of all faces of $C^\ddag$.
$C^{\ddag0}=\{\gamma \in \mathbb{R}^3\dvtx \gamma'\bolds{\theta} \leq
0, \forall  \bolds{\theta} \in C^\ddag\}$ is defined to be a polar
cone such that $(C^{\ddag0})^0=C^\ddag$. Following Shapiro (\citeyear{S1985}),
we can select a face $\nu\in\mathcal{F}(C^\ddag)$ corresponding to
the polar face $\nu^{0}\in\mathcal{F}(C^{\ddag0})$ such that the
linear spaces generated by $\nu$ and $\nu^0$ are orthogonal to
each other. For one face $\nu$ (or $\nu^0$), a projection $T_\nu$
(or $T_{\nu^0}$) [a symmetric idempotent matrix giving projection
onto the space generated by $\nu$ (or $\nu^0$)] can be found such
that $T_{\nu}=I-T_{\nu_0}$ since they are orthogonal. Then
$T_\nu\mathbf{Y}$ (or $T_{\nu^0}\mathbf{Y}$) is a projection of $\mathbf{Y}$ onto $C^\ddag$ (or $C^{\ddag0}$).

For a given $\mathbf{Y}$, let $g(\mathbf{Y}$) be the minimizer to
achieve the infimum in (\ref{A3}). Define
$\psi_{\nu|\mathbf{Y}}=\{\mathbf{Y} \in \mathbb{R}^3\dvtx g(\mathbf{Y})\in
\nu \}$ so that $g(\mathbf{Y})\in \nu$ if and only if $T_\nu
\mathbf{Y}\in C^\ddag$ and $T_{\nu0} \mathbf{Y}\in C^{\ddag0}$. By
Shapiro (\citeyear{S1985}), $g(\mathbf{Y})= T_{\nu}\mathbf{Y} \in C^\ddag$,
$\forall  \mathbf{Y} \in \psi_{\nu|\mathbf{Y}}$.

Note that the set $\psi_{\nu|\mathbf{Y}}$ is composed of $2^3$ disjoint
sets in $\mathbb{R}^3$. All these disjoint sets can be classified into
four categories as follows:
\begin{itemize}[(2)]
\item [(1)] $\psi_{\nu|\mathbf{Y}}^1=\{\mathbf{Y}; Y_1>0,Y_2>0,Y_3>0, g(\mathbf{Y})\in
\nu\}$,

\item [(2)] $\psi_{\nu|\mathbf{Y}}^2=\{\mathbf{Y}; Y_1>0,Y_2>0,Y_3\leq0, g(\mathbf{Y})\in \nu\}$; $\psi_{\nu|\mathbf{Y}}^3=\{\mathbf{Y}; Y_1>0,Y_2\leq0,Y_3>0, g(\mathbf{Y})\in \nu\}$; $\psi_{\nu|\mathbf{Y}}^4=\{\mathbf{Y}; Y_1\leq0,Y_2>0,Y_3>0, g(\mathbf{Y})\in
\nu\}$,

\item [(3)] $\psi_{\nu|\mathbf{Y}}^5=\{\mathbf{Y}; Y_1\leq0,Y_2\leq0,Y_3>0, g(\mathbf{Y})\in \nu\}$; $\psi_{\nu|\mathbf{Y}}^6=\{\mathbf{Y}; Y_1>0,Y_2\leq0,Y_3\leq0, g(\mathbf{Y})\in \nu\}$; $\psi_{\nu|\mathbf{Y}}^7=\{\mathbf{Y}; Y_1\leq0,Y_2>0,Y_3\leq0, g(\mathbf{Y})\in
\nu\}$,

\item [(4)] $\psi_{\nu|\mathbf{Y}}^8=\{\mathbf{Y}; Y_1\leq0,Y_2\leq0,Y_3\leq0, g(\mathbf{Y})\in \nu\}$.
\end{itemize}

By linear transformation, we cab define $C^\ast=\{\bolds{\theta}^\ast\dvtx \bolds{\theta}^\ast=\Lambda^{1/2}P'\bolds{\theta},
\forall \bolds{\theta} \in C^\ddag\}$ which is a polyhedral closed
convex cone. Then (\ref{A3}) can be further expressed as
%e12 ###
\begin{equation}\label{A4}
LR^*=\inf_{\bolds{\theta}^\ast \in C^\ast}\|\mathbf{z}-\bolds{\theta}^\ast\|^2,
\end{equation}
where $\mathbf{z}=\Lambda^{1/2}P'\mathbf{Y}$ [$P\Lambda P^T=I(\bolds{\theta}_0)$]
has a multivariate normal distribution with mean $\mathbf{0}$ and identity covariance matrix.

Let $C^{\ast0}$ be a polar cone of $C^{\ast}$ and
$(C^{\ast0})^0=C^\ast$. Two faces $\nu^\ast$ and $\nu^{\ast0}$ can
be defined with respect to $\mathcal{F}(C^\ast)$ and
$\mathcal{F}(C^{\ast0})$. The relevant orthogonal projections
$T_{\nu^\ast}$ and $T_{\nu^{\ast0}}$ corresponding to $\nu^\ast$
and $\nu^{\ast0}$ can be defined. Suppose $h(\mathbf{z}$) is the
minimizer to achieve the infimum in (\ref{A4}). Following
Shapiro (\citeyear{S1985}), a set $\psi_{\nu^\ast|\mathbf{z}}$ can be defined
similarly as $\psi_{\nu|\mathbf{Y}}$, such that $h(\mathbf{z})=
T_{\nu^\ast}\mathbf{z} \in C^\ast$, $\forall  \mathbf{z}$ $\in
\psi_{\nu^\ast|\mathbf{z}}$. It satisfies the conditions of Lemma 3.1
[Shapiro (\citeyear{S1985})]. Then we have
%e13 ###
\begin{equation}\label{A5}
\qquad LR^*= \|\mathbf{z}-h(\mathbf{z})\|^2=\|\mathbf{z}-T_{\nu^\ast}\mathbf{z}\|^2=\mathbf{z}'(I-T_{\nu^\ast})\mathbf{z}=\mathbf{z}'T_{\nu^{\ast0}}\mathbf{z}
\qquad \forall\mathbf{z}\in\psi_{\nu^\ast|\mathbf{z}}.
\end{equation}

Thus, the distribution of $LR^*$ in (\ref{A3}) can be evaluated by\vspace*{-1pt}
\begin{eqnarray*}%\label{A6}
&&\Pr(LR^\ast > c^2)\\[-2.5pt]
&&\qquad= \Pr\Biggl( \bigl(\mathbf{Y}-g(\mathbf{Y})\bigr)'I(\bolds{\theta}_0)\bigl(\mathbf{Y}-g(\mathbf{Y})\bigr)>c^2, \mathbf{Y}\in
\bigcup_{i=1}^{2^3}\psi_{\nu|\mathbf{Y}}^i\Biggr)                           \\[-2.5pt]
&&\qquad=\sum_{i=1}^{2^3} \Pr(\mathbf{Y}\in \psi_{\nu|\mathbf{Y}}^i) \Pr\bigl(\bigl(\mathbf{Y}-g(\mathbf{Y})\bigr)'I(\bolds{\theta}_0)\bigl(\mathbf{Y}-g(\mathbf{Y})\bigr)>c^2 |\mathbf{Y}\in \psi_{\nu|\mathbf{Y}}^i \bigr)        \\[-2.5pt]
&&\qquad=\sum_{i=1}^{2^3} \Pr(\mathbf{Y}\in \psi_{\nu|\mathbf{Y}}^i) \Pr(\mathbf{z}'T_{\nu^{\ast0}}\mathbf{z}>c^{2} |\mathbf{z}\in \psi_{\nu^\ast|\mathbf{z}}^i ),
%=&\sum_{i=1}^{2^3} Pr(\mathbf{Y}\in \psi_{\nu|\mathbf{Y}}^i) Pr(\chi_{rank(T_{\nu^{\ast0}})}^2>c^{2})\\
\end{eqnarray*}
where, conditional on $\mathbf{z} \in \psi_{\nu^\ast|\mathbf{z}}^i$,
$\mathbf{z}'T_{\nu^{\ast0}}\mathbf{z}$ is a chi-square distribution
[Lemma 3.1, Shapiro (\citeyear{S1985})]. By Bayes' theorem, the distribution of
$LR^*$ follows a mixture chi-square distribution with mixing
proportions $\Pr(\mathbf{Y}\in \psi_{\nu|\mathbf{Y}}^i)$ ($i=1,\ldots,2^3$)
and $\sum_{i=1}^{2^3} \Pr(\mathbf{Y}\in \psi_{\nu|\mathbf{Y}}^i)=1$.

The calculation of the mixture proportions follows Plackett (\citeyear{P1954}).
Specifically, when $\mathbf{Y}\in \psi_{\nu|\mathbf{Y}}^1$, $LR^*\sim
\chi_3^2$, and the corresponding mixture proportion
$w_3=\Pr(\mathbf{Y}\in\psi_{\nu|\mathbf{Y}}^1)
=\frac{1}{4\pi}[2\pi-\cos^{-1}\rho_{12}-\cos^{-1}\rho_{13}-\cos^{-1}\rho_{23}]$.
For category (2), $LR^*\sim \chi_2^2$ for $\mathbf{Y}\in
\psi_{\nu|\mathbf{Y}}^i$, $i=2,3,4$, with the corresponding mixture
probability calculated by
$w_2=\sum_{j=2}^4\Pr(\mathbf{Y}\in\psi_{\nu|\mathbf{Y}}^i)=\frac{1}{4\pi}[3\pi-\cos^{-1}\rho_{12|3}-\cos^{-1}\rho_{13|2}-\cos^{-1}\rho_{23|1}]$.
Correspondingly, $LR^*\sim \chi_1^2$ for $\mathbf{Y}\in
\psi_{\nu|\mathbf{Y}}^i$, $i=5,6,7$, with the relevant mixture
probability evaluated as
$w_1=\sum_{j=5}^7\Pr(\mathbf{Y}\in\psi_{\nu|\mathbf{Y}}^i)=\frac{1}{2}-w_3$
in category (3). For the last category, $LR^*\sim \chi_0^2$ for
$\mathbf{Y}\in \psi_{\nu|\mathbf{Y}}^8$ with the mixture probability
$w_0=\Pr(\mathbf{Y}\in\psi_{\nu|\mathbf{Y}}^8)=\frac{1}{2}-w_2$. Note
$\rho_{ab}$ is the correlation between the terms $a$ and $b$ calculated
from the Fisher information matrix, and
$\rho_{ab|c}=\frac{(\rho_{ab}-\rho_{ac}\rho_{bc})}{(1-\rho_{ac}^2)^{1/2}(1-\rho_{bc}^2)^{1/2}}$.
For more details of the derivation, the readers are referred to Li and
Cui (\citeyear{LC2009b}).
\end{appendix}

\section*{Acknowledgments}
We thank B. Larkins for providing the endosperm mapping data. We also
thank the Editor and two anonymous reviewers for helpful comments.
%This work was supported by NSF grant DMS-0707031 and by Michigan State
%University intramural research grant 06-IRGP-789.

\begin{supplement} [id=suppA]
\sname{Simulation and real data analysis}
\slink[doi]{10.1214/09-AOAS323SUPP}
\slink[url]{http://lib.stat.cmu.edu/aoas/323/supplement.zip}
\sdatatype{.zip}
\sdescription{Details for simulation are included in the supplemental
file. We also analyze the data with a Mendelian model. A comparison of
results with both imprinting and Mendelian models is summarized in the
supplemental file.}
\end{supplement}

\printaddresses

\end{document}